\documentclass[conference,letterpaper]{IEEEtran}
\IEEEoverridecommandlockouts
\usepackage{cite}
\usepackage{amsmath,amssymb,amsfonts}
\usepackage{algorithm}
\usepackage{algorithmic}
\usepackage{graphicx}
\usepackage{textcomp}
\usepackage{xcolor}
\usepackage{url}
\usepackage{amssymb, graphicx, amsmath, amsthm}
\usepackage{tikz}
\usepackage{multicol, blindtext}
\usepackage{subfigure}

\newtheorem{rk}{Remark}

\def\BibTeX{{\rm B\kern-.05em{\sc i\kern-.025em b}\kern-.08em
    T\kern-.1667em\lower.7ex\hbox{E}\kern-.125emX}}
\begin{document}
\title{A Guide to Reducing Carbon Emissions through Data Center Geographical Load Shifting\\

\thanks{The authors gratefully acknowledge support from the National Science Foundation under award CMMI-1832208.}
}

\author{\IEEEauthorblockN{Julia Lindberg, Yasmine Abdennadher, Jiaqi Chen, Bernard C. Lesieutre, Line Roald}
%\email{jrlindberg, abdennadher, jiaqi.chen, lesieutre, roald@wisc.edu}
\IEEEauthorblockA{Department of Electrical and Computer Engineering,
University of Wisconsin-Madison, Madison, WI 53706 USA}}
% <-this % stops an unwanted space
%\thanks{Corresponding author: J. Lindberg, jrlindberg@wisc.edu.}}

% The paper headers
%\markboth{Journal of \LaTeX\ Class Files,~Vol.~14, No.~8, August~2015}%
%{Shell \MakeLowercase{\textit{et al.}}: Bare Demo of IEEEtran.cls for IEEE Transactions on Magnetics Journals}
\maketitle
\begin{abstract}
Recent computing needs have lead technology companies to develop large scale, highly optimized data centers. These data centers represent large loads on electric power networks which have the unique flexibility to shift load both geographically and temporally. This paper focuses on how data centers can use their geographic load flexibility to reduce carbon emissions through clever interactions with electricity markets. Because electricity market clearing accounts for congestion and power flow physics in the electric grid, the carbon emissions associated with electricity use varies between (potentially geographically close) locations. Using our knowledge about this process, we propose a new and improved metric to guide geographic load shifting, which we refer to as the locational marginal carbon emission $\lambda_{\text{CO}_2}$. We compare this and three other shifting metrics on their ability to reduce carbon emissions and generation costs throughout the course of a year. Our analysis demonstrates that $\lambda_{\text{CO}_2}$ is more effective in reducing carbon emissions than more commonly proposed metrics that do not account for the specifics of the power grid.
\end{abstract}

%\keywords{data centers, load shifting, computing, carbon emissions}

\maketitle

\section{Introduction}
The recent technology revolution has led to an increase in demand for computing resources. Between 2010 and 2018 there was an estimated 550\% increase globally in the number of data center workloads and computing instances \cite{masanet2020recalibrating}. Technology companies like Amazon, Facebook, Google, Microsoft and Alibaba run networks of these highly optimized and efficient \emph{hyper-scale} data centers dispersed geographically throughout the world \cite{nrdc_2014, yevgeniy_analysts}.

Hyper-scale data centers represent large loads on electric power networks. In an effort to mitigate their environmental impact, many of the companies that operate these data centers have made public pledges to reduce their carbon emissions through improved efficiency and by investing in renewable power generation \cite{googleenvironmentalreport, amazonenvironment}. Google is working to become carbon free through the use of carbon-intelligent computing which shifts computing tasks to less carbon intensive hours or locations \cite{googleblog}. The concept of computing that adapts to the operation of the electric grid has been realized by start-ups \cite{lancium-press}, and plays an important part in the vision of zero carbon cloud computing \cite{yang2016zccloud, chien2019zero, deng2014harnessing}. 

Computing companies that operate networks of data centers have the ability to defer when computing tasks are processed or process them at different locations. This provides data centers with a unique tool to control and adapt their electricity use geographically. Previous research has examined the impact of integrating data centers and demand response \cite{dcdr_survey,liu2013data,liu2014pricing, zhou2016bilateral, zhou2020a, zhou2015when, chen2019an} or have considered geographical load shifting to reduce electricity costs \cite{rao_liu_xie_liu_2010, li_bao_li_2015, rao_liu_ilic_liu_2012, dou2017carbon}. The electricity markets are operated by independent system operators (ISOs) to minimize generation cost without direct consideration of carbon emissions. Other work has investigated the potential benefit of cooperation between data centers and the ISO \cite{rao_liu_ilic_liu_2012}, and modelled data center flexibility through the use of virtual links in time and space \cite{zhang2019flexibility}. 
Many of these works and others consider shifting of computing load to reduce the carbon emissions of data centers by increasing absorption of renewable energy \cite{goiri2013parasol, liu2012renewable, dou2017carbon, kim_yang_zavala_chien_2017, zheng2020mitigating}.
An important aspect of these previous works is that they either assume collaboration between the ISO and the data centers (meaning that data center operators must give up control of their energy usage to the market clearing), or perform load shifting using very simplified metrics of carbon emissions associated with the electric grid (i.e., they neglect variations in CO$_2$ emissions across different locations which arise due to transmission congestion). Here, we use additional data from the electric market clearing to compute the \emph{locational marginal carbon emissions}, a metric that more accurately represents the carbon emissions associated with electricity usage at different locations in the grid.

In electricity markets the price of electricity is calculated based on locational marginal prices (LMPs), which reflect the increase in system cost for one additional unit of load. Previous work on reducing carbon emissions through load shifting has assumed that prices are directly tied to the fraction of non-renewable energy \cite{2015liu}, or considered average carbon emissions for electricity in a region and/or renewable energy curtailment \cite{chien2015zero, yang2017large, zheng2020mitigating}. A benefit to these metrics is that several companies provide information about the average carbon intensity of electricity \cite{tomorrow, caisoemissions} or total renewable energy curtailment \cite{caisocurtailment}, which makes the metrics easier to compute. However, these metrics fail to consider important aspects of electric grid operation, such as the impact of a marginal increase or decrease in load or transmission capacity.

The use of marginal emissions as a tool to assess the impact of interventions such as energy efficiency or renewable energy has been discussed in \cite{siler2012marginal, callaway2018location, 2010Ruiz, 2011Rudkevich,tabors2021methodology}. Recently, a measure of locational marginal carbon emissions was proposed for data center geographical load shifting in \cite{lindberg2020the}. The challenge of using locational marginal carbon emissions is that the computation of this metric relies on information from the market clearing regarding system topology as well as price and emissions information from individual generators. Thus, while \cite{lindberg2020the} proposed load shifting based on locational marginal carbon emissions and provided some initial results on the efficacy of this metric, the paper did not provide a comprehensive assessment of how this load shifting metric compares with other metrics or how it would perform over a longer period of time. The purpose of this paper is to examine the different carbon emissions metrics and get a better understanding of their accuracy, efficacy and overall behavior. 

We structure our results as follows: In Section~\ref{sec:model} we review two models for data center-driven load shifting, one where the data centers shift load independent of coordination with the ISO, and one where the ISO controls the data center flexibility. In Section~\ref{carbonmeasures} we outline four different metrics to guide data center-driven load shifting as well as two new metrics to evaluate the efficacy of the load shifting. Section~\ref{testcase} provides an overview of our test case, while Section~\ref{sec:comparisonofshiftingmetrics} compares the carbon savings achieved with different models and shifting metrics.

\section{Models for load shifting} \label{sec:model}
Electricity markets are typically cleared using a DC optimal power flow (DC OPF) model, which minimizes generation cost subject to transmission and generation constraints \cite{christie2000transmission,litvinov2010}. 
We distinguish between two different modes of interaction between the data centers and the ISO that operates the electricity market.
\begin{itemize}
    \item[(1)]\textbf{Data center-driven load shifting.} In this model, the data center loads are fixed values to the ISO, but can be adapted by the data centers themselves. Here the data centers solve an internal optimization problem (reflecting internal constraints on the load flexibility) to determine how to shift their electricity consumption to reduce carbon emissions. 
    \item[(2)]\textbf{ISO-driven load shifting.} In this model, the data center interacts with the market as a market participant with flexible demand, and provides a model of load flexibility to the ISO. The ISO then integrates the load flexibility model into the overall market clearing, and determines how to shift data center load to achieve the best solution for the whole system. 
\end{itemize}
The goal of this paper is to understand how the carbon emission reductions achieved through data center-driven load shifting compare for different load shifting metrics, and how large of emission reductions the data center-driven load shifting can achieve relative to the ISO-driven load shifting. We outline both models below. 

\subsection{Data center-controlled load shifting}\label{sec:datacenterdrivenloadshifting}
We first describe the data center driven load shifting model, which is adopted from \cite{lindberg2020the}. This model operates in a three stages.

\vspace{1mm}
\noindent \textbf{Step 1: The ISO solves a DC OPF.} In the first step, the ISO clears the electricity market by solving the DC OPF. To formulate the DC OPF, we consider an electric power network with the set of nodes, loads, transmission lines and generators denoted $\mathcal{N}$, $\mathcal{D}$, $\mathcal{L}$ and $\mathcal{G}$ respectively. Let $\mathcal{G}_i \subset \mathcal{G}$ and $\mathcal{D}_i \subset \mathcal{D}$ be the subset of generators and loads connected to node $i$. Given this notation, the DC OPF is defined as:
\begin{subequations}\label{dcopf}
\begin{align}
\min_{\theta, P_g} ~~&c^T P_g \label{dcopfcost} \\
\text{s.t.} ~~& \textstyle \sum_{\ell\in\mathcal{G}_i}  \!P_{g,\ell} -\! \textstyle \sum_{\ell\in\mathcal{D}_i} P_{d,\ell} = &&
\nonumber\\
&\qquad\textstyle\sum_{j:(i,j)\in\mathcal{L}} \!\!\!\!-\beta_{ij}(\theta_i \!- \!\theta_j), &&\forall i\in\mathcal{N} \label{balance}\\
-&P^{\text{lim}}_{ij} \!\leq\! -\beta_{ij}(\theta_i \!-\!\theta_j) \!\leq\! P^{\text{lim}}_{ij}, &&\forall (i,j)\in\mathcal{L}
\label{lineineq}\\
& P^{\min}_{g,i} \leq P_{g,i} \leq P^{\max}_{g,i}, && \forall i\in\mathcal{G}
\label{genineq}\\
& \theta_{\text{ref}} = 0. \label{refnode} 
\end{align}
\end{subequations}
Here, the optimization variables are the voltage angles at each node, $\theta_i$ for $i \in \mathcal{N}$ as well as the generation dispatch $P_{g, \ell}$ for all $\ell \in \mathcal{G}$. The objective value \eqref{dcopfcost} seeks to minimize generation costs where $c \in \mathbb{R}^{|\mathcal{G}|}$ is a vector of generator costs and $P_g$ is the vector of all generator variables $P_{g, \ell}$. The constraint \eqref{balance} ensures that nodal power balance constraints are met, where $\beta_{ij} \in \mathbb{R}$ is the susceptance value on line $(i,j)$ and $P_{d,\ell}$ is the load demand at load $\ell \in \mathcal{D}_i$. The constraints \eqref{genineq} and \eqref{lineineq} define transmission line and generator capacity constraints where $P_{ij}^{\text{lim}}$ is the transmission capacity, which we assume is the same in both directions, and $P_{g}^{\min}$ and $P_{g}^{\max}$ are the generator capacity constraints.  Finally, \eqref{refnode} fixes the voltage angle at the reference node to be zero. 
\vspace{1mm}

\noindent \textbf{Step 2: Data centers shift load.} Independently of the ISO, data center operators shift their load to minimize carbon emissions. To estimate the impact of a load shift, the data centers utilize a shifting metric $\lambda$. There are multiple possible definitions of $\lambda$, which will be discussed in Section~\ref{sec:different shifting metrics}. We let $\mathcal{C}$ denote the set of all shiftable data center loads and consider optimization variables $\Delta P_{d,i}$ for all $ i \in \mathcal{C}$ and $s_{ij}$ for all $(i,j) \in \mathcal{C} \times \mathcal{C}$.  The former represents the change in load at data center $i$ and the latter represents the shift in load from data center $i$ to $j$. The resulting optimization problem is given by:
\begin{subequations}\label{datacenteropt}
\begin{align}
\min_{\Delta P_{d}, s} \ \  & \sum_{i \in \mathcal{C}} \lambda_{i} \Delta P_{d,i} + \sum_{(i,j) \in \mathcal{C} \times \mathcal{C}} d_{ij} s_{ij} \label{loadshiftobj} \\
 \text{s.t. } \ \   &\Delta P_{d,i} = \textstyle\sum_{j\in\mathcal{C}} s_{ji} - \textstyle\sum_{k\in\mathcal{C}} s_{ik} \quad &&\forall i\in{C} \label{datacenterflex1} \\
    \textstyle &\sum_{i\in\mathcal{C}} \Delta P_{d,i}= 0 \label{datacenterflex2} \\
    &- \epsilon_i \cdot P_{d,i} \leq \Delta P_{d,i} \leq \epsilon_i \cdot P_{d,i} \quad &&\forall i\in\mathcal{C} \label{datacenterlim1} \\
   & 0 \leq s_{ij} \leq M_{ij} \quad &&\forall ij\in\mathcal{C}\times\mathcal{C}. \label{datacenterlim2} 
\end{align}{}
\end{subequations}
The objective value \eqref{loadshiftobj} minimizes $\lambda$ over the shift in data center load, $\Delta P_d$, while considering a cost $d_{ij}$ associated with shifting load from data center $i$ to data center $j$. The constraint \eqref{datacenterflex1} defines that the change in load at data center $i$ is equal to the total load shifted in minus the total load shifted out, \eqref{datacenterflex2} enforces the sum of all load shifts to be zero, \eqref{datacenterlim1} limits the amount each data center can shift as a percentage, $\epsilon$, of their original load and \eqref{datacenterlim2} limits how much load data center $i$ can send to data center $j$.
\vspace{1mm}

\noindent \textbf{Step 3: ISO resolves DC OPF with new load pattern.} Next, the ISO resolves the DC OPF \eqref{dcopf} with new load profile, $P_{d,i}' = P_{d,i} + \Delta P_{d,i}^*$, where $\Delta P_{d,i}^*$ is the optimal solution to \eqref{datacenteropt} for all $i \in \mathcal{N}$. We assume that the system is operated using this solution\footnote{Note that we assume that the ISO solves the OPF twice for each time step, once before and once after the load shift. In reality, OPF is only solved once for each time step, and the data center loads would likely use $\lambda$ values calculated based on the OPF solution from the previous time step for shifting. However, if the OPF model is solved frequently enough, e.g. every 5 minutes, it is reasonable to assume that the load will remain largely constant between time periods and our model is a good estimate.}

\subsection{Shifting Metrics}\label{sec:different shifting metrics}

For data centers to shift load as described in \eqref{datacenteropt}, the shifting metric $\lambda$ needs to be specified. Below, we review four different metrics that have been proposed to guide load shifting. Note that all the metrics are defined as vectors with one entry for each node in the system. 
\vspace{1mm}

\noindent \textbf{The price of electricity $\lambda_{\text{LMP}}$.} The most widely used metric for load shifting is the price of electricity, which is given by the locational marginal price (LMP) at each node in the power grid. We will refer to this metric as $\lambda_{\text{LMP}}$. The LMP represents the increase in overall system cost due to an incremental increase of $1$ MW of load at the given node, and is calculated as the dual variable of the nodal balance constraints \eqref{balance} of the original DC OPF. LMPs are easy to access, as they are typically made available in real time, and have been proposed for data center load shifting in \cite{rao_liu_ilic_liu_2012}. Furthermore, since renewable generators tend to be the cheaper generators, it is often assumed that shifting load to nodes with lower prices $\lambda_{\text{LMP}}$ will contribute to reducing renewable energy curtailment.
\vspace{1mm}

\noindent \textbf{Average carbon emissions $\lambda_{\text{average}}$.} A common metric for the carbon content of electricity is average carbon emissions per MW of load across a region of the electric grid \cite{tomorrow}. This metric, which we will denote by $\lambda_{\text{average}}$, has the same value for all nodes in a region $\mathcal{R}$. The $k$th entry of $\lambda_{\text{average}}$, corresponding to the $k$th node located in region $\mathcal{R}$, is defined as
\begin{align}
 \textstyle \lambda_{\text{average}} = \frac{\sum_{i \in \mathcal{R}} g_i \cdot P_{g,i}  }{ \sum_{i\in \mathcal{R}} P_{g,i}},
\end{align}
where $g_i$ is the carbon intensity of generator $i$. The intuition behind this metric is to shift load to regions with lower average carbon footprint, and thus reducing the average carbon emissions associated with electricity consumption. This type of metric has been proposed for data center load shifting in \cite{zheng2020mitigating}. A benefit of using $\lambda_{\text{average}}$ is that these values are made publicly available by various companies \cite{tomorrow, caisoemissions}.
\vspace{1mm}

\noindent \textbf{Excess low carbon power $\lambda_{\text{excess}}$.} This metric considers shifting based on the amount of excess low carbon generation capacity available in a region, and accounts for renewable energy curtailment (solar PV and wind) as well as unused hydro, nuclear power and storage generation. The excess low carbon power is defined for a given region, where the value of the $k$th component of $\lambda_{\text{excess}}$ is the same for all nodes $k$ in a region $\mathcal{R}$.
Let $e_i$ be the excess capacity (MW) for each low carbon generator $P_{g,i}$ (with $e_i=0$ for other generators). The $k$th component in $\lambda_{\text{excess}}$ is given by
\begin{align}
    \lambda_{\text{excess}} = - \sum_{i \in \mathcal{R}} e_i.
\end{align}
This metric provides incentive to shift load to regions with high amounts of excess low carbon power (i.e., more negative values of $\lambda_{\text{excess}}$), which could allow for more utilization of low carbon generation. This metric has been proposed for inter-regional data center load shifting in \cite{zheng2020mitigating}.
\vspace{1mm}

\noindent \textbf{Locational marginal carbon emissions $\lambda_{\text{CO}_2}$.} 
This metric is defined as the change in carbon emission as a function of the change in load at a given node $k$ in the network. Similar to the $\lambda_{\text{LMP}}$, it is derived by considering how the DC OPF solution would change given a change in the load at a specific location. However, instead of considering the change in the objective function (which measures overall system cost), the derivation of $\lambda_{\text{CO}_2}$ uses sensitivity analysis of linear programs to identify the change in the carbon emissions. The following derivation is adopted from \cite{lindberg2020the}.

Consider an optimal solution $x^* = [\theta^*, P_g^*] \in \mathbb{R}^n$ to the DC OPF \eqref{dcopf}. From linear optimization theory we know there exists at least one basic optimal solution with $Ax^* = b$, where $A \in \mathbb{R}^{n \times n}$ is a full rank matrix consisting of the coefficients for all the binding constraints of \eqref{dcopf} at the optimal solution $x^*$ and $b$ is the right hand side. Specifically, the rows of $A$ consist of the equality constraints \eqref{balance} and \eqref{refnode} as well as a subset of the inequality constraints \eqref{lineineq}, \eqref{genineq} that are satisfied at equality for $x^*$.

A small change in load can be represented as a small change in the right hand side $b$, given by $\Delta b = \begin{bmatrix} \Delta P_d & 0 \end{bmatrix}^T$. Assuming that the change is sufficiently small to not alter the set of active constraints, we can compute the associated change in generation as $A\Delta x = \Delta b$ where $\Delta x = [\Delta \theta \ \Delta P_g]$, giving the linear relationship

\begin{align}
    \begin{bmatrix} \Delta \theta \\ \Delta P_g   \end{bmatrix} &= A^{-1} \cdot \begin{bmatrix} \Delta P_{d} \\ 0 \end{bmatrix}
\end{align}

Denote the matrix consisting of the last $|\mathcal{G}|$ rows and first $|\mathcal{N}|$ columns of $A^{-1}$ by $B$. This gives the linear relationship between load and generation changes, $\Delta P_g = B \cdot  \Delta P_d$. 

Let $g \in \mathbb{R}^{| \mathcal{G}|}$ be a cost vector that measures the carbon emissions of each generator per MW. The $i$th component of $g$, $g_i$, is the carbon intensity of generator $i$. Multiplying each side of $\Delta P_g = B \cdot  \Delta P_d$ on the left by $g$ gives us the following carbon sensitivity:

\begin{align}
    \Delta CO_2 = g \cdot \Delta P_g = g \cdot B \cdot \Delta P_d = \lambda_{\text{CO}_2}  \Delta P_d \label{newobj}.
\end{align}
where we define $\lambda_{\text{CO}_2} = g \cdot B$. Intuitively we think of the $k$th component of $\lambda_{\text{CO}_2}$ as measuring how an increase of $1$ MW of load at node $k$ will affect the total carbon emissions of the system.
 
The benefit of $\lambda_{\text{CO}_2}$ is that it captures the impact of load shifting on carbon emissions in a more direct way than $\lambda_{\text{LMP}}$ and includes detailed information regarding the marginal impact of load shifting at specific data center locations compared with $\lambda_{\text{average}}$ and $\lambda_{\text{excess}}$. 
However, since the full network information needed to calculate the $\lambda_{\text{CO}_2}$ is not made publicly available, it is not easy to calculate these values in real time. In addition, since $\lambda_{\text{CO}_2}$ represents a linear sensitivity, these values may only be accurate for small load shifts.

\begin{rk}\label{remark1}
Marginal carbon emissions at a node depends on the objective function considered when calculating an operating point. In a situation where the ISO minimizes carbon emissions instead of generation costs by replacing the objective function \eqref{dcopfcost} with $g^T P_g$ where $g \in \mathbb{R}^{|\mathcal{G}|}$ is as defined above, the values of $\lambda_{\text{CO}_2}$ could be obtained as the dual variables to the nodal power balance constraints in the same way that $\lambda_{\text{LMP}}$ is obtained for the DC OPF where the ISO is minimizing cost.
\end{rk}

\subsection{ISO-controlled load shifting}\label{sec:isodrivendatacenterloadshifting}

The second model for load shifting considered in this paper assumes that the data centers give their flexibility to the ISO, which uses this flexibility to achieve overall system objectives. We primarily consider this situation as a benchmark to the load shifting model defined in Section~\ref{sec:datacenterdrivenloadshifting}.
\vspace{1mm}

\noindent \textbf{DC OPF with flexibility (DC OPF-FLEX).} In the DC OPF-FLEX model, the ISO considers the following optimization problem which includes data center load shifting flexibility:
\begin{subequations}\label{isocontrolshifting}
\begin{align}
    \min_{P_g, \theta, \Delta P_d, s} ~~&\  c^T P_g + \textstyle\sum_{ij} d_{ij} s_{ij} \label{isocontrolobj}\\
     \text{s.t.} ~~& \textstyle \sum_{\ell\in\mathcal{G}_i}  \!P_{g,\ell} -\! \textstyle \sum_{\ell\in\mathcal{D}_i} (P_{d,\ell} + \Delta P_{d,\ell}) = &&
\nonumber\\
&\qquad\textstyle\sum_{j:(i,j)\in\mathcal{L}} \!\!\!\!-\beta_{ij}(\theta_i \!- \!\theta_j), &&\forall i\in\mathcal{N} \label{isocontrolconst1}\\
&\text{Constraints } \eqref{lineineq}, \eqref{genineq}, \eqref{refnode} \label{isocontrol1}\\
&\text{Constraints } \eqref{datacenterflex1}, \eqref{datacenterflex2}, \eqref{datacenterlim1}, \eqref{datacenterlim2} \label{isocontrol2}
\end{align}
\end{subequations}
The optimization variables $\theta_i, i \in \mathcal{N}$ and $P_{g,\ell}, \ell \in \mathcal{G}$ are as in DC OPF \eqref{dcopf}. The variables $\Delta P_{d, \ell}$ for $\ell \in \mathcal{C}$ and $s_{ij}$ for $(i,j) \in \mathcal{C} \times \mathcal{C}$ are as defined in the data center driven shifting model \eqref{datacenteropt}. The objective value \eqref{isocontrolobj} minimizes the cost of electricity production where $c \in \mathbb{R}^{|\mathcal{G}|}$ is a vector of generator costs with the additional cost $d_{ij}$ of shifting load from data center $i$ to data center $j$. Constraints \eqref{isocontrolconst1}-\eqref{isocontrol2} ensure the nodal power balance, line limits, generation capacity, reference node and data center load shifting flexibility constraints are met.
\vspace{1mm}

\noindent \textbf{Cost of carbon emissions.} In the formulation of DC OPF \eqref{dcopf} and DC OPF-FLEX \eqref{isocontrolshifting}, we assumed that the ISO is minimizing the overall cost of operating the electric system, without consideration of carbon emissions. This is representative of system operations today. However, as mentioned in Remark~\ref{remark1}, it is also possible that the ISO may change their objective and include a cost on the carbon emissions. In this case, we consider objective function 
\begin{align}
    [ \alpha c^T + (1-\alpha) g^T ] \cdot P_g \label{carboncosttradeoff}
\end{align}
where $\alpha$ is a weighting factor that represents the emphasis on minimizing generation cost versus reducing carbon emissions. 
\vspace{1mm}

\noindent \textbf{Benchmarking.}  We use the DC OPF-FLEX model as a benchmark for two reasons. First, many demand response schemes allow flexible loads to participate in markets by providing information about their flexibility to the ISO. Thus, the DC OPF-FLEX model \eqref{isocontrolshifting} represents a realistic model of potential future interactions between data centers and the ISO. 
Second, all of the load shifting metrics that are proposed to guide data center load shifting in Section \ref{sec:different shifting metrics} provide only partial information about the impact of a load shift.  In comparison, DC OPF-FLEX is able to optimize the load shift with exact knowledge of how the cost and/or carbon emissions will change as a result. Therefore, we can expect that these models will always find the most optimal load shift, i.e., the load shift that gives the lowest cost or carbon emissions. Theoretical results comparing these models were derived in \cite{lindberg2020the}.

\section{Evaluation Metrics}\label{carbonmeasures}
Some evaluation metrics for data center efficiency and carbon emissions are provided in \cite{aravanis2015metrics}. These metrics assess the flexibility and sustainability of data centers as well as the potential benefit to upgrading data center equipment. In contrast, we want to evaluate how shifting load geographically affects total system generation costs and carbon emissions.

Since the data centers represent a small subset of loads and only a relatively small percentage of the load at each data center is allowed to shift, the percentage decrease or increase is frequently only a small fraction of the total system emissions and cost. For this reason, we introduce two metrics that measure the change in generation cost and carbon emissions relative to the (small) amount of load that is shifted. The main purpose for introducing these two metrics is to normalize the change in carbon emissions and generation cost by the small amount of load being shifted.
\vspace{1mm}

\noindent \textbf{Reduction per allowed MW: $\mathbf{\mu_{\text{percent, CO}_2}}$.} This metric is defined as the change in carbon emission normalized by the maximum amount of load that can be shifted. The maximum amount of load that can be shifted (in MW), $\mathbb{L}$, can be calculated as $\mathbb{L} = \sum_{i \in \mathcal{C}} \epsilon_i \cdot P_{d,i}$. Let $\Delta$ be the total change in carbon emissions from the original DC OPF. We then define the relative reduction $\mu_{\text{percent, CO}_2}$ as
\begin{align*}
    \mu_{\text{percent, CO}_2} &=   \frac{\Delta}{\mathbb{L}} .
\end{align*}
The units of $\mu_{\text{percent, CO}_2}$ are carbon tons per MW. This metric is a measure of how well a method is able to make use of the available flexibility. 
\vspace{1mm}

\noindent \textbf{Reduction per shifted MW: $\mathbf{\mu_{\text{shift, CO}_2}}$.} This metric is similar to $\mu_{\text{percent, CO}_2}$ but considers the change in carbon emissions when normalized by the actual load shift, as opposed to the maximum allowed load shift. Again, let $\Delta$ be the total change in carbon from the original DC OPF. Denote $\mathcal{S} = \sum_{i \in \mathcal{C}} | \Delta P_{d,i}| $ as the total amount of load shifted (in MW). We then define
\begin{align}
    \mu_{\text{shift, CO}_2} &= \frac{\Delta }{\mathcal{S}} .
\end{align}
The units of $\mu_{\text{shift, CO}_2}$ are carbon tons per MW. This metric is a measure of the change in carbon emissions per MW shifted. Notice that if in each time step we shift the maximal amount of load as dictated by $\epsilon$, then $\mu_{\text{shift, CO}_2} = \mu_{\text{percent, CO}_2}$. 

The definitions of $\mathbf{\mu_{\text{percent, CO}_2}}$ and $\mathbf{\mu_{\text{shift, CO}_2}}$ can easily be adapted to assess cost reductions by defining $\Delta$ as the total change in cost relative to the original DC OPF. We denote the corresponding cost reduction metrics as $\mathbf{\mu_{\text{percent, \$}}}$ and $\mathbf{\mu_{\text{shift, \$}}}$. 
\vspace{1mm}

\noindent \textbf{Predicted vs actual impact of load shifting.}
For the data center-driven load shifting, all evaluation metrics mentioned above can be defined either for the predicted impact of the load shift obtained by considering the objective function of the data center problem \eqref{datacenteropt} in Step 2 of our model, and for the actual load shift obtained after the ISO resolves the DC OPF \eqref{dcopf} in Step 3. By evaluating the difference between the two, we can assess the accuracy of the data center-driven load shifting and check whether this model is overly optimistic when predicting the impact of a load shift.

\section{Test Case}\label{testcase}

We next perform an extensive year long analysis of the different methods to guide data center load shifting, using the data center-driven shifting model and shifting metrics outlined in Section~\ref{sec:model}. Considering a full year of operations allows us to provide information based on a range of different operating conditions which gives a good overall idea of the behavior of each shifting metric. 

For our analysis we use the IEEE RTS-GMLC system \cite{barrows2020the}. This system has $73$ buses, $158$ generators and $120$ lines. The network has three regions, $\mathcal{R}_1 = \{v_1,\ldots,v_{24} \}$, $\mathcal{R}_2 = \{v_{25}, \ldots, v_{48} \}$ and $\mathcal{R}_3 = \{ v_{49}, \ldots, v_{73} \}$, which are used as definitions of the regions when evaluating $\lambda_{\text{average}}$ and $\lambda_{\text{excess}}$. Since the original system does not contain any loads that are designated as data centers, we assign data centers at buses $3,7,28$ and $70$. We assume that each of the data centers consume a fixed power of $250$ MW throughout the year. For all other loads and renewable generation, we use the hourly load and generation data provided with \cite{barrows2020the}. The system serves a total of $44,821,000$ MWh of load over the course of the whole year and the data centers account for $8,784,000$ MWh or $19.6 \%$ of the total energy consumption (though the relative share varies over time). 

Adding these large data center loads to the network greatly increases the total system load and results in time steps where the original DC OPF is infeasible. To remedy this we set the minimum generation constraint, $P_g^{\text{min}} = 0$ for all $g \in \mathcal{G}$, and increase the maximum generation constraints by $50 \%$ of the original value. We allow each data center to shift up to $20\%$ of its load, or $50$ MW, and put no limitations on how much load each data center can shift to one another. This means throughout the year at most $1,756,800$ MW of load can shift, or $3.92 \%$ of the total system load.

\section{Comparison of Shifting Metrics}\label{sec:comparisonofshiftingmetrics}

\begin{table*}[h!]
\centering
\begin{tabular}{ |c|c|c|c|c|c|c|c| } 
 \hline
  & CO$_2$ tons & Generation Cost $(\$)$ & Curtailment (MW)  & $\mu_{\text{percent, CO$_2$}}$ & $\mu_{\text{percent, $\$$}}$ & $\mu_{\text{shift, CO$_2$}}$ & $\mu_{\text{shift, $\$$}}$ \\ 
  \hline 
  DC OPF & $13,964,000$& $322,690,000$ & $4,010,100$ &  &   & & \\
    DC OPF-FLEX & $13,873,000$ $ (-0.65 \%)$ & $319,188,000$ $(-1.09 \%)$ & $3,895,707$ $(-2.85 \%)$ & $-0.05$ &$-2.00$ & $ - 0.06 $ & $- 2.31$\\
 $\lambda_{\text{CO}_2}$ & $13,763,000$ $(-1.44 \%)$ &$321,830,000$ $(-0.27 \%)$ & $3,908,700$ $(-2.52 \%)$ & $-0.11$ & $-0.49$ & $- 0.14 $ & $-0.62 $  \\ 
 $\lambda_{\text{LMP}}$ & $13,869,000$ $(-0.68\%)$ & $319,260,000 $ $(-1.07 \%)$ & $3,896,500$ $(-2.83 \%)$ & $ -0.05$ & $ -1.96 $ & $ - 0.06$ & $- 1.99 $\\ 
 $\lambda_{\text{average}}$ &  $13,870,000$ $(-0.68 \%)$ & $320,990,000$ $(-0.53 \%)$ &$3,916,700$ $( -2.33 \%)$ &  $-0.06 $ & $-0.95 $ & $-0.09 $ & $-1.64 $  \\
 $\lambda_{\text{excess}}$ & $13,983,000$  $(+ 0.14 \%)$ & $ 323,470,000$ $(+ 0.24 \%)$ & $4,056,600$ $(+1.16 \%)$ &  $ +0.01$ & $ +0.44$ &  $ + 0.03 $ & $ + 1.11$
\\
 \hline
\end{tabular}
\vspace{1mm}
\caption{Carbon emissions, generation cost and curtailment after shifting based on different shifting metrics. }
\label{actualchange}
\end{table*}

\begin{table*}[h!]
\centering
\begin{tabular}{ |c|c|c|c|c|c|c|c| } 
 \hline
  & CO$_2$ tons & Generation Cost $(\$)$ & Curtailment (MW)  & $\mu_{\text{percent, CO$_2$}}$ & $\mu_{\text{percent, $\$$}}$ & $\mu_{\text{shift, CO$_2$}}$ & $\mu_{\text{shift, $\$$}}$\\ 
  \hline 
  DC OPF & $13,964,000$& $322,690,000$ & $4,010,100$ &  & & &   \\ 
    DC OPF-FLEX & $13,873,000$ $ (-0.65 \%)$ & $319,188,000$ $(-1.09 \%)$ & $3,895,707$ $(-2.85 \%)$ & $-0.05$ &$-2.00$ & $ - 0.06 $ & $- 2.31$\\
 $\lambda_{\text{CO}_2}$ & $13,623,000$ $(-2.45\%)$ & $319,590,000$ $(-0.96 \%)$ & $3,812,000$ $(-4.94 \%)$  & $-0.19 $ &  $-1.77 $ & $-0.25 $ & $-2.22 $ \\ 
 $\lambda_{\text{LMP}}$ & $13,807,000$ $(-1.12 \%)$ & $316,370,000$ $(-1.96 \%)$ & $3,810,100$ $(-4.99 \%)$ &  $ -0.09$ &  $-3.60 $  & $ -0.09$ & $-3.66 $ \\ 
 $\lambda_{\text{average}}$ & $13,766,000$ $(-1.42 \%)$ & $318,760,000$ $(-1.22 \%)$ & $3,831,700$ $(-4.45 \%)$ & $-0.11 $ &  $ -2.22$ & $-0.20 $ & $ -3.83$\\
 $\lambda_{\text{excess}}$ & $13,926,000$ $(-0.27 \%)$ & $321,960,000$ $(-0.23 \%)$ & $4,003,400$ $(-0.17 \%)$ &  $ -0.02$ &  $-0.42 $ & $-0.05 $ & $-1.05 $\\
 \hline
\end{tabular}
\vspace{1mm}
\caption{Predicted carbon emissions, generation cost and curtailment after shifting based on the different shifting metrics.}
\label{predictedchange}
\end{table*}

In this section we compare the performance of the different shifting metrics and benchmark against ISO-controlled load shifting. 

\subsection{Carbon Emissions Reduction from ISO-controlled Load Flexibility}
\begin{figure}[h!]
 \centering
    \includegraphics[width = 0.4\textwidth]{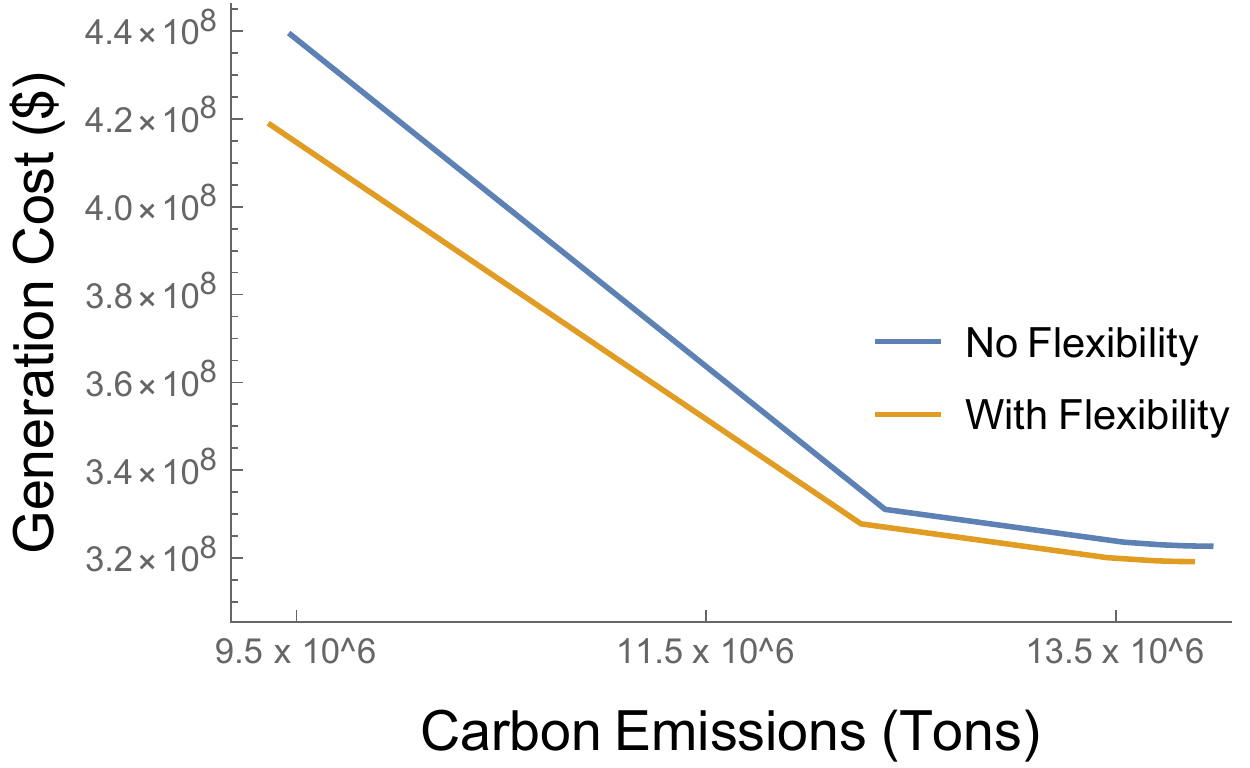}
    \caption{Trade off as ISO minimizes cost and carbon with and without data center flexibility.}
    \label{fig:ISOcontrol}
\end{figure}

We first compare the impact of ISO-controlled load shifting on generation cost and carbon emission reduction by comparing results obtained from the DC OPF \eqref{dcopf} and the DC OPF-FLEX \eqref{isocontrolshifting}. We repeatedly solve these two problems with cost function~\eqref{carboncosttradeoff} and values of $\alpha$ ranging from $0$ to $1$. Figure~\ref{fig:ISOcontrol} shows the generation cost and carbon emissions across all the different solutions for the cases with and without data center load flexibility. The low generation cost/high carbon emissions solution in the bottom right corner corresponds to the setting with $\alpha=1$, where the ISO only considers cost minimization in their solution. The high generation cost/low carbon in the top left corner corresponds to the case with $\alpha=0$, where the ISO minimizes carbon emissions. The intermediate values of alpha gives rise to the solutions along the blue and yellow lines. 

We observe that as $\alpha$ is reduced, there is first a large decrease in carbon emissions with only a small increase in generation cost, but as we approach $\alpha=0$, there is both a large decrease in carbon emissions and a large increase in generation cost. The point $\alpha = 0.1$ is where the gradient of each curve changes, indicating that if the ISO would even lightly weight minimizing carbon emissions, substantial reductions could be achieved without dramatically increasing generation costs. This trend is similar for both cases (with/without flexibility), but the solution with flexibility consistently has lower generation cost for a solution with comparable carbon emissions. Similarly, the solution with flexibility can achieve lower carbon emissions at similar generation cost. This demonstrates the benefits of flexibility in general.

However, in our case, we are interested in understanding the impact of flexibility on carbon emission reductions for a specific choice of cost function. We can gain an initial understanding of this by comparing the difference in carbon emissions for the solutions without and with flexibility for $\alpha=1$, i.e., when we only minimize cost. These two solutions correspond to the rightmost points on the blue and yellow lines. We observe that the solution with flexibility (yellow line) has lower carbon emissions than the solution with no flexibility (blue line). Specifically, the carbon emissions are reduced by 91 CO$_2$ tons (0.65\%). This indicates that for a given cost function, the carbon emissions reduction achieved through data center load flexibility only amounts to a small reduction of the total system emissions, even in the best case scenario where the ISO is using all available system information to make the best possible use of the additional flexibility.

\subsection{Comparison of Outcomes with Different Shifting Metrics}\label{sec:comparisonofoutcomes}
We next compare the performance of data center-controlled load shifting with different shifting metrics against each other and against the ISO-controlled benchmark.
Using the data center-driven load shifting model and network parameter values as described in Section~\ref{testcase},  we obtain solutions for the whole year with each shifting metrics. We also include the results obtained from the initial DC OPF (without load flexibility, minimizing cost) and from the DC OPF-FLEX (with load flexibility, minimizing cost). The results in Table~\ref{actualchange} represent the final results of the load shifting \emph{after} the ISO has solved the DC OPF with the shifted load. The two first lines represent the benchmark results with the DC OPF and the DC OPF-FLEX, and the shifting metrics are ranked based on their total achieved carbon emission reduction. Observe that the DC OPF and DC OPF-FLEX models are both evaluated when $\alpha = 1$.

We notice that the $\lambda_{\text{CO}_2}$ metric is by far the best in terms of reducing carbon emissions, both in terms of the total carbon emission reduction and the reduction per shifted MW of load ${\mu_{\text{shift, CO}_2}}$. On average, for every $1$ MW shifted, we save $0.14$ tons of carbon. The total carbon emission reduction is  $-1.44\%$ which is twice as large as the reduction achieved with the DC OPF-FLEX. This is a particularly interesting result, because it indicates that data center-driven load shifting with respect to $\lambda_{\text{CO}_2}$ achieves higher carbon reductions than if the data centers were to relinquish their flexibility to the ISO. This supports the idea that direct action by the data centers to minimize carbon emissions can make a substantial difference in carbon emission reductions. We note that the carbon emission reduction is achieved while the system cost is reduced by $0.27\%$.

Next, we observe that shifting with respect to  $\lambda_{\text{LMP}}$ results in carbon and cost savings that are comparable to the case where the ISO assumes control of the data center flexibility. The two cases decrease carbon emissions by $-0.68\%$ and $-0.65\%$ and give an overall decrease in generation cost of $1.07 \%$ and $1.09 \%$, respectively. This similarity is explained by observing that when the data centers shift with respect to $\lambda_{\text{LMP}}$ they have similar objectives and the same constraints as when the ISO assumes control of the data center flexibility. The main difference is that DC OPF-FLEX gives a global solution while shifting with respect to $\lambda_{\text{LMP}}$ relies on local sensitivity information.

Further, we observe that shifting with respect to $\lambda_{\text{average}}$ gives a similar overall decrease in carbon as $\lambda_{\text{LMP}}$. However, when the reduction is considered per MW shifted as in $\mu_{\text{shift}}$, we see that $\lambda_{\text{average}}$ results in a greater carbon savings per MW shifted. This demonstrates that if a cost were to be associated to shifting load, $\lambda_{\text{average}}$ would be more effective.

Finally, the shifting metric based on excess availability of low carbon power $\lambda_{\text{excess}}$ actually increases the carbon emissions and generation costs of the network. This is a counter-intuitive since shifting load to a region of the network with the most excess low carbon resources seems like it should allow the data centers to make more use of low carbon generation sources. This is clearly not always the case as the actual location of the data center must be such that the data center can access the low carbon power. Line constraints within a region can prevent a data center from having access to these low carbon resources.

\subsection{Comparison of Accuracy}
For each of the data center-driven shifting metrics, we can assess the accuracy of the metric as the difference between the predicted carbon emission reduction (obtained from the data center internal optimization problem \eqref{datacenteropt} in Step 2) and the actual change in carbon emissions (computed from based on the DC OPF in Step 3). 

Table~\ref{predictedchange} and Table~\ref{actualchange} show the predicted and actual change in carbon emissions, generation cost and curtailment, respectively. By comparing the results in the two tables, we see that the predicted carbon savings are better than the actual savings for all of the metrics. This implies that the shifting metrics are not able to provide an entirely accurate picture of the impact of load shifting. This is largely because the shifting model relies on a linear approximation of \eqref{dcopf}, which assumes the binding constraints remain the same before and after shifting. This need not be the case, which then makes the original linearization inaccurate.

In particular we notice that $\lambda_{\text{excess}}$ predicts an overall decrease of $14,000$ tons of carbon or $-0.05$ tons per MW shifted, but actually produces a solution which leads to an increase in carbon emissions as seen in Table~\ref{actualchange}. Further we observe that $\lambda_{\text{average}}$ predicts a better carbon savings then $\lambda_{\text{LMP}}$, both overall and per MW shifted. However, this contradicts the results in Section~\ref{sec:comparisonofoutcomes} which indicated that $\lambda_{\text{LMP}}$ lead to a better overall savings.

\section{Conclusion}\label{conclusion}

This paper extended previous work \cite{lindberg2020the} introducing a model for technology companies to exploit the geographic load shifting flexibility of data centers to reduce carbon emissions independently of collaboration with an ISO. We reviewed four shifting metrics and two new evaluation metrics and performed an extensive analysis of these metrics on a years worth of data. We found that $\lambda_{\text{CO}_2}$ outperformed all other metrics in terms of achieving carbon emission reductions. Importantly, the carbon savings achieved with this metric were larger than when the data center provided load flexibility directly to the ISO (who use this flexibility to minimize cost). 

This work demonstrates the effectiveness of shifting with respect to $\lambda_{\text{CO}_2}$ motivating the question of how to estimate these values in real time. Since these values are not currently made publicly available, future work largely involves a better understanding of how $\lambda_{\text{CO}_2}$ varies as a function of other available system values. 

\section{Acknowledgements}
We acknowledge support from the National Science Foundation under award CMMI-1832208.

\bibliographystyle{unsrt}
\bibliography{bibfile}

\end{document}